\documentclass[12pt,a4paper,reqno]{article}
\usepackage{amsmath}

\advance\textheight\topskip
\allowdisplaybreaks

\renewcommand{\i}{\mathrm{i}}
\newcommand{\e}{\mathrm{e}}
\newcommand{\al}{\alpha}
\newcommand{\ad}{{\dot{\alpha}}}
\newcommand{\be}{\beta}
\newcommand{\de}{\delta}
\newcommand{\ep}{\varepsilon}
\newcommand{\bpsi}{\bar{\psi}}
\newcommand{\si}{\sigma}
\newcommand{\bsi}{\bar{\sigma}}
\newcommand{\cD}{\mathcal{D}}
\newcommand{\cF}{\mathcal{F}}
\newcommand{\cH}{\mathcal{H}}
\newcommand{\half}{\tfrac{1}{2}}
\newcommand{\tab}{\quad\,}
\newcommand{\p}{\partial}
\newcommand{\bD}{\bar{D}}
\newcommand{\com}[2]{[\,#1\, ,\,#2\,]}
\newcommand{\frc}[2]{\frac{\raisebox{-2pt}{$#1$}}{#2}}
\DeclareMathOperator{\im}{Im}

\DeclareSymbolFont{AMSb}{U}{msb}{m}{n}
\DeclareMathSymbol{\fieldR}{\mathalpha}{AMSb}{"52}

\begin{document} 

\begin{flushright} \small
 ITP-UH-06/00 \\ hep-th/0005044
\end{flushright}
\medskip

\begin{center}
 {\large\bfseries New N=2 Supersymmetric Vector-Tensor Interaction}
 \\[5mm]
 Ulrich Theis
 \\[2mm]
 {\small\slshape
 Institut f\"ur Theoretische Physik, Universit\"at Hannover, \\
 Appelstra\ss{}e 2, 30167 Hannover, Germany \\[2pt]
 utheis@itp.uni-hannover.de}
\end{center}
\vspace{5mm}

\hrule\bigskip

\centerline{\bfseries Abstract} \medskip
An $N=2$ supersymmetric self-interaction of the vector-tensor multiplet
is presented, in which the vector provides the gauge field for local
central charge transformations. The dual description in terms of a
vector multiplet and an $N=1$ superspace formulation are given.
\bigskip

\hrule\vspace{5mm}


Since the discovery of the relevance of the $N=2$ vector-tensor
multiplet \cite{SSW} to certain heterotic string theory
compactifications \cite{dWKLL}, its possible interactions have been
investigated in a number of publications \cite{CdWFKST,CdWFT,GHH}. In
\cite{CdWFKST,DT} the central charge of the multiplet was gauged by a
coupling to an abelian vector multiplet that provides the gauge field
for local central charge transformations, while in \cite{CdWFT} a
self-interaction of the multiplet was introduced, giving rise to a
Chern-Simons coupling of the vector and tensor. These two kinds of
interactions have subsequently been rederived in harmonic superspace
from deformations of the superfield constraints that determine the
multiplet \cite{DK,DIKST}.

In the present letter it is shown how the central charge of the
vector-tensor multiplet can be gauged by a self-interaction, without a
coupling to an additional vector multiplet. The model is obtained from a
deformation of the free component action and its corresponding
supersymmetry and gauge transformations by means of the Noether method.
This results in a realization of the supersymmetry algebra which holds
only on-shell, i.e.\ on fields satisfying their equations of motion.
We shall address the problem of constructing a manifestly $N=2$
supersymmetric off-shell formulation in a future publication and content
ourselves with an off-shell version in terms of $N=1$ superfields for
the time being.

Gauge field theories of the kind considered here have been investigated
in detail in \cite{BT} in the framework of $N=1$ supersymmetry. They are
actually members of a family of (bosonic) models discovered by Henneaux
and Knaepen in \cite{HK}, that in four spacetime dimensions involve
non-polynomial interactions of 1- and 2-forms. In \cite{BT} it was shown
that every four-dimensional Henneaux-Knaepen model admits an $N=1$
supersymmetric generalization. The vector-tensor multiplet with gauged
central charge as found in \cite{CdWFKST} was so far the only known
model with two supersymmetries. In the following we present the second
example.
\medskip

The vector-tensor multiplet contains a 1-form $A_\mu$, a 2-form
$B_{\mu\nu}$, a real scalar $\phi$, and an SU(2) doublet of Weyl
spinors $\psi_\al^i$. The bosonic and fermionic off-shell degrees of
freedom can be matched by inclusion of a real auxiliary scalar $U$.

The supersymmetry algebra of the free multiplet involves a central
charge that acts non-trivially on the gauge fields,
 \begin{equation}
  \Delta_c A_\mu = c\, H_\mu\ ,\quad \Delta_c B_{\mu\nu} = c *\!
  F_{\mu\nu}\ ,\quad c \in \fieldR\ .
 \end{equation}
Here $*F^{\mu\nu}=\ep^{\mu\nu\rho\si}\p_\rho A_\si$ and $H^\mu=\half
\ep^{\mu\nu\rho\si}\p_\nu B_{\rho\si}$ are the Hodge-dual field
strengths of $A_\mu$ and $B_{\mu\nu}$ respectively. The conserved
current that corresponds to this global symmetry of the free action is
simply
 \begin{equation}
  J^\mu = F^{\mu\nu} H_\nu\ .
 \end{equation}
A coupling of $A_\mu$ to this current yields a consistent interaction
vertex $gA_\mu J^\mu$ of dimension five (where accordingly $g$ is a
coupling constant of dimension $-1$). It requires to modify the gauge
transformations of $A_\mu$ and $B_{\mu\nu}$ by first-order terms that
are precisely the central charge transformations given above,
 \begin{align}
  \de A_\mu & = \p_\mu C(x) + g C(x) H_\mu + O(g^2) \notag \\[2pt]
  \de B_{\mu\nu} & = \p_\mu C_\nu(x) - \p_\nu C_\mu(x) + g C(x) *\!
	F_{\mu\nu} + O(g^2)\ ,
 \end{align}
which amounts to gauging the global symmetry. In order to restore
supersymmetry to order $g$, the vertex $A_\mu J^\mu$ is to be
accompanied by additional terms of dimension five, which however depend
on the gauge potentials only via the field strengths and therefore do
not introduce further first-order deformations of the gauge
transformations. The question now is whether the action and the
transformations can be extended such that both gauge invariance and
$N=2$ supersymmetry are realized to all orders in the coupling constant.

Since the interaction falls into the family of Henneaux-Knaepen models,
it is to be expected that the resulting action and transformations are
non-polynomial in the fields, so a Noether construction order by order
seems impractical. As it turns out, however, the introduction of an
auxiliary vector $V^\mu$, which to lowest order in $g$ equals $H^\mu$
on-shell, greatly simplifies the construction: it essentially suffices
to determine the deformation up to second order, all higher-order
contributions are due to prefactors that are functions of the scalar
field $\phi$ only, which are easily completed to all orders.

The full Lagrangian constructed in this way reads
 \begin{align}
  \mathcal{L} & = \frc{1}{2} \cos(2g\phi)\, V^\mu V_\mu - V_\mu H^\mu
	- \frc{1}{4} \cos(2g\phi)\, \cF^{\mu\nu} \cF_{\mu\nu} -
	\frc{1}{4} \sin(2g\phi) *\! \cF^{\mu\nu} \cF_{\mu\nu} \notag \\
  & \tab + \frc{1}{2} \cos(2g\phi)\, \p^\mu \phi\, \p_\mu \phi - \i
	\cos(2g\phi)\, \big( \psi^i \si^\mu \overset{\leftrightarrow}{
	\p_\mu} \bpsi_i \big) + 2g \sin(2g\phi)\, V_\mu\, \psi^i \si^\mu
	\bpsi_i + \frc{1}{2}\, U^2 \notag \\
  & \tab - \i g\, \cF_{\mu\nu} \big( \e^{\i g\phi} \psi^i \si^{\mu\nu}
	\psi_i - \e^{-\i g\phi} \bpsi^i \bsi^{\mu\nu} \bpsi_i \big)
	- \frc{g^2}{\cos(2g\phi)}\, \big( \e^{-\i g\phi} \psi^i \psi^j
	+ \e^{\i g\phi} \bpsi^i \bpsi^j \big)^2\ . \label{L}
 \end{align}
Here $\cF_{\mu\nu}$ is an extended field strength,
 \begin{equation}
  \cF_{\mu\nu} = (\p_\mu + g V_\mu) A_\nu - (\p_\nu + g V_\nu) A_\mu\ .
 \end{equation}

The Lagrangian is invariant, up to a total derivative, under gauge
transformations
 \begin{align}
  \de A_\mu & = (\p_\mu + g V_\mu)\, C \notag \\[2pt]
  \de B_{\mu\nu} & = g C \big[ \cos(2g\phi) *\! \cF_{\mu\nu} -
	\sin(2g\phi)\, \cF_{\mu\nu} - 2 g \big( \e^{\i g\phi} \psi^i
	\si_{\mu\nu} \psi_i + \e^{-\i g\phi} \bpsi^i \bsi_{\mu\nu}
	\bpsi_i \big) \big] \notag \\
  & \tab + \p_\mu C_\nu - \p_\nu C_\mu \notag \\[2pt]
  \de \phi & = \de \psi_\al^i = \de V_\mu = \de U = 0\ , \label{trafo}
 \end{align}
and rigid supersymmetry transformations generated by
 \begin{align}
  \cD_\al^i \phi & = \psi^i_\al \notag \\[2pt]
  \cD_\al^i V_\mu & = - \i \p_\mu \psi^i_\al \notag \\[2pt]
  \cD_\al^i A_\mu & = \i g\, \psi^i_\al A_\mu + \i\, \e^{\i g\phi}
	(\si_\mu \bpsi^i)_\al \notag \\[2pt]
  \cD_\al^i B_{\mu\nu} & = 2 \cos(2g\phi)\, (\si_{\mu\nu} \psi^i)_\al
	- 2 g\, \e^{-\i g\phi} A_{[\mu} (\si_{\nu]} \bpsi^i)_\al
	\notag \\[2pt]
  \cD_\al^i \bpsi^j_\ad & = \half \ep^{ij} \si^\mu_{\al\ad} (V_\mu 
	+ \i \p_\mu \phi) \notag \\[2pt]
  \cD_\al^i \psi^{\be j} & = \half \ep^{ij}\, \e^{\i g\phi}
	\cF_{\mu\nu}\, \si^{\mu\nu}{}_{\!\al}{}^\be + \tfrac{\i}{2}
	\ep^{ij} \de^\be_\al\, U - \i g\, \ep^{ij} \psi^k_\al
	\psi^\be_k - g \tan(2g\phi)\, \de^\be_\al\, \psi^i \psi^j
	\notag \\
  & \tab - \frc{\i g}{\cos(2g\phi)}\, \de^\be_\al\, \bpsi^i \bpsi^j
	\notag \\
  \cD_\al^i U & = - \cos(2g\phi)\, (\si^\mu \p_\mu \bpsi^i)_\al + g
	\sin(2g\phi)\, (\p_\mu \phi - \i V_\mu)\, (\si^\mu \bpsi^i
	)_\al \notag \\
  & \tab - g\, \e^{\i g\phi} \cF_{\mu\nu} (\si^{\mu\nu} \psi^i)_\al
	- \frc{2\i\, g^2}{\cos(2g\phi)}\, \big( \e^{-2\i g\phi} \psi^i
	\psi^j + \bpsi^i \bpsi^j \big)\, \psi_{\al j}\ .
 \end{align}
Using the transformation properties of $\cF_{\mu\nu}$,
 \begin{align}
  \de \cF_{\mu\nu} & = g C\, (\p_\mu V_\nu - \p_\nu V_\mu) \notag
	\\[2pt]
  \cD_\al^i \cF_{\mu\nu} & = \i g\, \psi^i_\al \cF_{\mu\nu} + 2\i\,
	(\p_{[\mu} + g V_{[\mu})\, \big( \e^{\i g\phi} \si_{\nu]}
	\bpsi^i \big)_\al\ , \label{trafoF}
 \end{align}
gauge invariance is easily verified, while supersymmetry requires
some effort.

Note that although the combination $\p_\mu+gV_\mu$ resembles a covariant
derivative, there is no gauge transformation associated with $V_\mu$,
it is merely an auxiliary field. In order to eliminate it, we collect
all terms involving $V_\mu$. They can be written as
 \begin{equation}
  \mathcal{L} = \frc{1}{2}\, V_\mu K^{\mu\nu} V_\nu - V_\mu \cH^\mu
  + \dots\ ,
 \end{equation}
where we have employed the abbreviations
 \begin{align}
  K^{\mu\nu} & = \cos(2g\phi)\, \big[ \eta^{\mu\nu} (1 - g^2 A \cdot\!
	A) + g^2 A^\mu A^\nu \big] \notag \\[2pt]
  \cH^\mu & = \half \ep^{\mu\nu\rho\si} \p_\nu B_{\rho\si} + g \cos(2g
	\phi)\, F^{\mu\nu}\! A_\nu + g \sin(2g\phi) *\! F^{\mu
	\nu}\! A_\nu \notag \\
  & \tab - 2 g \sin(2g\phi)\, \psi^i \si^\mu \bpsi_i + 2\i\, g^2 \big(
	\e^{\i g\phi} \psi^i \si^{\mu\nu} \psi_i - \e^{-\i g\phi}
	\bpsi^i \bsi^{\mu\nu} \bpsi_i \big) A_\nu\ .
 \end{align}
The equation of motion of $V_\mu$ then yields ($\approx$ denotes
on-shell equality)
 \begin{equation}
  V_\mu \approx (K^{-1})_{\mu\nu} \cH^\nu\ ,\quad (K^{-1})_{\mu\nu}
  = \frc{\eta_{\mu\nu} - g^2 A_\mu A_\nu}{\cos(2g\phi)\, (1 - g^2 A
  \cdot\! A)}\ ,
 \end{equation}
which upon insertion into the Lagrangian and the transformations results
in expressions non-polynomial in both $gA_\mu$ and $g\phi$. If one
expands the Lagrangian in powers of $g$, the coupling $gA_\mu J^\mu$ is
recovered to first order, which was the goal of the construction.
\medskip

The supersymmetry algebra closes on-shell; if the equations of motion
hold, the commutator of two supersymmetry transformations $\Delta_\xi=
\xi^\al_i\cD_\al^i+\bar{\xi}_\ad^i\bar{\cD}_i^\ad$ yields a translation
and a gauge transformation,
 \begin{equation}
  \com{\Delta_\xi}{\Delta_\zeta} \approx a^\mu \p_\mu - \de\ ,
 \end{equation}
with parameters
 \begin{align}
  a^\mu & = \i \big( \zeta_i \si^\mu \bar{\xi}^i - \xi_i \si^\mu
	\bar{\zeta}^i \big) \notag \\[2pt]
  C & = a^\mu\! A_\mu + \frc{\i}{\raisebox{+2pt}{$g$}} \big( \e^{\i g
	\phi} \xi_i \zeta^i + \e^{-\i g\phi} \bar{\xi}^i \bar{\zeta}_i
	\big) \notag \\[2pt]
  C_\mu & = a^\nu B_{\nu\mu} + \frc{1}{2g} \sin(2g\phi)\, a_\mu +
	\big( \e^{\i g\phi} \bar{\xi}^i \bar{\zeta}_i - \e^{-\i g\phi}
	\xi_i \zeta^i \big) A_\mu\ .
 \end{align}
Note that while the parameter $C$ is singular for $g=0$, the
transformation $\de$ is not, since $C$ occurs either under a derivative
or with a factor $g$.

It is quite remarkable that the gauged central charge transformations
do not commute with supersymmetry transformations, not even on-shell.
Rather, one has
 \begin{equation}
  \com{\Delta_\xi}{\de} \approx \de'\ ,
 \end{equation}
where now
 \begin{align}
  C' & = \i g\, C \big( \xi_i \psi^i - \bar{\xi}^i \bpsi_i \big) \notag
	\\[2pt]
  C_\mu' & = g C \big( \e^{-\i g\phi} \xi^i \si_\mu \bpsi_i - \e^{\i g
  \phi} \psi^i \si_\mu \bar{\xi}_i \big)\ .
 \end{align}
\medskip


A free vector-tensor multiplet is dual to an abelian vector multiplet.
Since in our interacting model the 2-form $B_{\mu\nu}$ occurs only via
its field strength, it too can be dualized into a pseudo-scalar. In
fact, the use of an auxiliary vector makes the dualization particularly
simple. Considering $H^\mu$ as an independent field and implementing
the Bianchi identity $\p_\mu H^\mu=0$ by means of a Lagrange multiplier
field $\varphi$, we find that according to the equation of motion of
$H^\mu$, we can replace $V_\mu$ with $\p_\mu \varphi$,
 \begin{equation}
  V_\mu \rightarrow \p_\mu \varphi\ ,\quad \cD_\al^i \varphi = - \i
  \psi_\al^i\ .
 \end{equation}
$\phi$ and $\varphi$ can be combined into a complex scalar $X$ that is
chiral, $\bar{\cD}_{\ad i}X=0$. With suitable field redefintions, we
obtain the (on-shell) field content of an $N=2$ vector multiplet with
the standard supersymmetry transformations\footnote{Every function of
$(\varphi-\i\phi)$ is chiral, but only $X$ as in \eqref{Vmult} satisfies
the additional constraint $\cD^i\cD^jX+\bar{\cD}^i\bar{\cD}^j\bar{X}=0$
that gives rise to a vector multiplet.},
 \begin{equation} \label{Vmult}
  X = \frc{1}{2g}\, \e^{g(\varphi-\i\phi)}\ ,\quad \hat{A}_\mu =
  \e^{g\varphi}\! A_\mu\ ,\quad \lambda_\al^i = \e^{g(\varphi-\i\phi)}
  \psi_\al^i\ .
 \end{equation}
Up to a factor, $\cF_{\mu\nu}$ turns out to be nothing but the field
strength of the new vector $\hat{A}_\mu$,
 \begin{equation}
  \cF_{\mu\nu} = \e^{-g\varphi} \hat{F}_{\mu\nu}\ .
 \end{equation}
The holomorphic prepotential $f(X)$ that determines the action of the
vector multiplet can be obtained from the field strength terms in the
Lagrangian \eqref{L},
 \begin{equation*}
  - \frc{1}{4} \cos(2g\phi)\, \cF^{\mu\nu} \cF_{\mu\nu} - \frc{1}{4}
  \sin(2g\phi)\, *\! \cF^{\mu\nu} \cF_{\mu\nu} = - \frc{1}{4} \im
  \Big[ \frc{\i}{(2gX)^2}\, \hat{F}_{\mu\nu} (\hat{F}^{\mu\nu} - \i
  *\! \hat{F}^{\mu\nu}) \Big]\ . \notag
 \end{equation*}
Comparing with the general expression common to every vector multiplet,
 \begin{equation*}
  - \frc{1}{4} \im \big[ f''(X)\, \hat{F}_{\mu\nu} (\hat{F}^{\mu\nu}
  - \i *\! \hat{F}^{\mu\nu}) \big]\ ,
 \end{equation*}
we read off the second derivative of the prepotential, and conclude that
 \begin{equation}
  f(X) = - \frc{\i}{4g^2}\, \ln (gX)\ .
 \end{equation}
\medskip


Finally, we briefly demonstrate how the present model can be derived
from an $N=1$ superspace integral, drawing on results obtained in
\cite{BT}:

Let us embed $B_{\mu\nu}$, $\phi$ and $\psi_\al^1$ in a chiral spinor
superfield $\Psi_\al(x,\theta,\bar{\theta})$, $A_\mu$ and $\psi_\al^2$
in a real superfield $A(x,\theta,\bar{\theta})$, and $V_\mu$ in a real
superfield $V(x,\theta,\bar{\theta})$ with a mass dimension shifted by
$+1$. From these, we construct field strength superfields $Y_\al(x,
\theta,\bar{\theta})$ and $W_\al(x,\theta,\bar{\theta})$,
 \begin{equation}
  Y_\al = - \frc{\i}{4}\, \bD^2 \big[ \e^{-2\i gV} D_\al (\e^{\i gV}\!
  A) \big]\ ,\quad W_\al = \frc{1}{2} \bD^2 D_\al V\ ,
 \end{equation}
where $D_\al$ and $\bD_\ad$ are the usual $N=1$ supercovariant
derivatives. $\Psi_\al$, $Y_\al$ and $W_\al$ are chiral,
 \begin{equation}
  \bD_\ad \Psi_\al = \bD_\ad Y_\al = \bD_\ad W_\al = 0\ ,
 \end{equation}
so the action
 \begin{equation}
  S = \int\! d^4 x\, d^2 \theta\, \Big( W^\al \Psi_\al + Y^\al Y_\al
  - \frc{1}{4g^2}\, d^2 \bar{\theta}\ \e^{2\i gV} \Big) + \text{c.c.}
 \end{equation}
is manifestly $N=1$ supersymmetric. The component gauge transformations
\eqref{trafo} follow from the superfield transformations
 \begin{equation}
  \de \Psi_\al = \i \bD^2 D_\al R - 2\i g\, \Lambda Y_\al\ ,\quad
  \de A = \i \big( \e^{\i gV}\! \Lambda - \e^{-\i gV}\! \bar{\Lambda}
  \big)\ ,\quad \de V = 0\ ,
 \end{equation}
where $R(x,\theta,\bar{\theta})$ is a real and $\Lambda(x,\theta,
\bar{\theta})$ a chiral superfield. Gauge invariance of the action is
then due to the relations
 \begin{equation}
  \de Y_\al = \i g\, \Lambda W_\al\ ,\quad D^\al W_\al - \bD_\ad
  \bar{W}^\ad = 0\ ,
 \end{equation}
the former being the superfield analog of the gauge transformation
\eqref{trafoF} of $\cF_{\mu\nu}$. Passing to Wess-Zumino gauge for $A$
and $\Psi_\al$ and eliminating the auxiliary fields (except $V_\mu$)
results in the $N=2$ supersymmetric Lagrangian \eqref{L}.

Evidently, there is an infinite number of $N=1$ supersymmetric
Henneaux-Knaepen models that describe the field content of a
vector-tensor multiplet, for one may specify different functions of
$V$ (the $d^2\theta\,d^2\bar{\theta}$-part in the action) that allow
to eliminate the auxiliary fields (another example was given in
\cite{BT}), but in general these will not possess a second
supersymmetry.
\medskip

The previously known interactions of the vector-tensor multiplet can
all be derived by dimensional reduction from interactions of the
$N=(1,0)$ tensor multiplet in six dimensions \cite{BSvP,DIKST}. In
these cases the gauged central charge is a remnant of translations in
the additional spacelike directions. It would be interesting to know
whether this applies to the present model as well, or whether the fact
that the gauge field for the local central charge transformations
resides in the vector-tensor multiplet itself requires a different
mechanism in six dimensions. Models that might be relevant in this
context can be found in \cite{NS}.

The new model introduced here indicates that our current knowledge about
the possible interactions of the vector-tensor multiplet is far from
being exhaustive. In particular, the problem of finding interactions
between several such multiplets has not been attacked successfully as
yet. A possible starting point would be to consider the general $N=1$
supersymmetric Henneaux-Knaepen models of \cite{BT} and to single out
those that exhibit an SU(2) $R$-symmetry. Work in this direction is
under way.
\medskip

\emph{Remark:} In the course of this work, a further $N=2$
supersymmetric Henneaux-Knaepen model was found, involving two vector
multiplets and a double-tensor multiplet \cite{B}.
\bigskip

\textbf{Acknowledgements} \\[6pt]
I would like to thank Friedemann Brandt and Sergei Kuzenko for helpful
discussions and suggestions. Thanks also to Augusto Sagnotti for
bringing the work on six-dimensional tensor multiplet couplings to my
attention.

\end{document}